\documentclass{article}

\usepackage{arxiv}
\usepackage[utf8]{inputenc} 
\usepackage[T1]{fontenc}
\usepackage{graphicx} 

\usepackage[T1]{fontenc} 
\usepackage{amsmath}
\usepackage[cmintegrals]{newtxmath}
\usepackage{bm} 
\usepackage{csquotes}
\usepackage{multirow}
\usepackage{tabularx}
\usepackage{booktabs}
\usepackage{multicol}
\usepackage{listings}
\usepackage{textgreek}
\usepackage{footnote}
\usepackage{comment}
\usepackage{url}
\usepackage{caption}
\usepackage[sort&compress,numbers]{natbib}
\usepackage{subcaption}
\title{Lessons learned from replicating a study on information-retrieval based test case prioritization}
\author{Nasir Mehmood Minhas\\  Department of Software Engineering, Blekinge Institute of Technology \\nasir.mehmood.minhas@bth.se \AND Mohsin Irshad \\ Ericsson Sweden AB, Karlskrona, Sweden \\ mohsin.irshad@ericsson.com \AND Kai Petersen \\ Department of Software Engineering, Blekinge Institute of Technology \\University of Applied Sciences Flensburg, Germany \\ kai.petersen@bth.se \AND J\"{u}rgen B\"{o}rstler \\ Department of Software Engineering, Blekinge Institute of Technology \\ jurgen.borster@bth.se}

\renewcommand{\shorttitle}{Minhas et al.}
\date{}

\begin{document}

\maketitle


 
 \begin{abstract} 

\textbf{Objective}: In this study, we aim to replicate an artefact-based study on software testing to address the gap. We focus on (a) providing a step by step guide of the replication, reflecting on challenges when replicating artefact-based testing research, (b) Evaluating the replicated study concerning its validity and robustness of the findings. \\
\textbf{Method}: We replicate a test case prioritization technique by Kwon et al. We replicated the original study using four programs, two from the original study and two new programs. The replication study was implemented using Python to support future replications.\\
\textbf{Results}: Various general factors facilitating replications are identified, such as: (1) the importance of documentation; (2) the need of assistance from the original authors; (3) issues in the maintenance of open source repositories (e.g., concerning needed software dependencies); (4) availability of scripts. We also raised several observations specific to the study and its context, such as insights from using different mutation tools and strategies for mutant generation.\\
\textbf{Conclusion}: We conclude that the study by Kwon et al. is replicable for small and medium programs and could be automated to facilitate software practitioners, given the availability of required information.
 \end{abstract}

\section{Introduction}
\label{Sec:intro}

Replications help in evaluating the results, limitations, and validity of studies in different contexts \cite{shull2008role}. They also help establishing or expanding the boundaries of a theory \cite{da2014replication,shull2008role}.

During the previous four decades, software engineering researchers have built new knowledge and proposed new solutions. Many of these lack consolidation \cite{krein2010case} replication studies can help in establishing the solutions and expanding the knowledge. Software engineering researchers have been working on replication studies since the 1990s. Still, the number of replicated studies is small, and a more neglected area is the replication of software testing experiments \cite{cruz2019replication,santos2021comparing,da2014replication,krein2010case}. Most software engineering replication studies are conducted for experiments involving human participants; few replications exist for artefact-based experiments \cite{da2014replication}. 
 
In the artefacts-based software engineering experiments, the majority of the authors use the artefacts from the software infrastructure repository (SIR) \cite{yoo2012regression}. Do et al.\cite{do2005supporting} introduced SIR in 2005 to facilitate experimentation and evaluation of testing techniques and to promote replication of experiments and aggregation of findings.

Researchers proposed different techniques to support regression testing practice, and there are various industrial evaluations of regression testing techniques. Adopting these techniques in practice is challenging because the results are inaccessible for the practitioners \cite{bin2019search}. Replications of existing solutions on regression testing can be helpful in this regard, provided the availability of data and automation scripts of these replications.

Attempts have been made to replicate regression testing techniques. The majority of these replications are done by the same group of authors who originally proposed the techniques \cite{do2004empirical,do2006use, do2010effects}. There is a need for conducting more independent replications in software engineering \cite{do2005supporting}. However, evidence of independent replications in regression testing is low \cite{da2014replication}.

Overall, we would highlight the following research gaps concerning replications:
 
\begin{itemize}
\item \emph{Gap 1: Only a small portion of studies are replications: } Among the reasons for a lower number of replications in software engineering is the lack of standardized concepts, terminologies, and guidelines \cite{krein2010case}. Software engineering researchers need to make an effort to replicate more studies. 
 
\item \emph{Gap 2: Lack of replication guidelines:} There is a need to work on the guidelines and methodologies to support replicating the studies \cite{de2015investigations}.
 
\item \emph{Gap 3: Lack of replications in specific subject areas:} Software testing as a subject area has been highlighted as an area lacking replication studies \cite{da2014replication}. According to Da Silva et al. \cite{da2014replication} the majority of replication studies focuses on software construction and software requirements. Despite a well-researched area, the number of replication studies in software testing is at the lowest than other software engineering research areas according to Magalh{\~a}es et al. \cite{de2015investigations}.
 
\item \emph{Gap 4: Lack of studies on artefact-based investigations:} Only a few replicated studies focused on artefact-based investigations \cite{da2014replication}. That is, the majority of studies focused on experiments and case studies involving human subjects. Artefact-based replications are of a particular interest as they require to build and run scripts for data collection (e.g., solution implementation and logging), and at the same time compile and run the software systems, which are the subject of study. 
\end{itemize}
Considering the gaps stated above, we formulate the following research goal: 
 
\begin{center}
        \framebox{
       \parbox[t][1.6cm]{12cm}{
       
        \vspace{0.1mm}
       \textit{\textbf{Goal:} To replicate an artefact-based study in the area of software testing, with a focus on reflecting on the replication process and the ability to replicate the findings of the original study.}
      
         } 
        }
\end{center} 

\vspace{0.2 cm}

To achieve our research goal, we present the results of our replication of an IR-based test case prioritization technique proposed by Kwon et al. \cite{kwon2014test}. The authors introduced a linear regression model to prioritize the test cases targeting infrequently tested code. The inputs for the model are calculated using term frequency (TF), inverse document frequency (IDF), and code coverage information \cite{kwon2014test}. TF and IDF are the weighing scheme used with information retrieval methods \cite{roelleke2013information}.  
The original study's authors used open-source data sets (including SIR artefacts) to evaluate the proposed technique. We attempted to evaluate the technique using four programs to see if the replication confirms the original study's findings. We selected two programs from the original study and two new cases to test the technique's applicability on different programs. 

Our research goal is achieved through the following:

\begin{enumerate}
\item \textit{Objective 1: Studying the extent to which the technique is replicable.} Studying the extent to which the technique is replicable and documenting the detail of all steps will help draw valuable lessons. Hence, contributing with guidance for future artefact-based replications (Gap 2, Gap 4).
\item \textit{Objective 2: Evaluating the results of the original study \cite{kwon2014test}.}  Evaluating the results through the replication provides an assessment of the validity and the robustness of the results of the original study. Overall, we contribute to the generally limited number of replication studies (Gap 1) in general, and replication studies focused on software testing in particular (Gap 3). 
\end{enumerate}

The organization of the rest of the paper is as follows:
Section \ref{Sec:BG} provides a brief introduction to the concepts relevant to this study. Section \ref{Sec:RW} presents a brief discussion of some replications carried out for test case prioritization techniques. Along with the research questions and summary of the concepts used in the original study, Section \ref{Sec:Method} describes the methodologies we have used to select the original study and conduct the replication. Threats to the validity of the replication experiment are discussed in Section \ref{Sec:valid}. 
Section \ref{Sec:Res} presents the findings of this study, Section \ref{Sec:Disc} provides the discussion on the findings of replication study, and Section \ref{Sec:con} concludes the study.

\section{Background}
\label{Sec:BG}
This section provides a discussion on the topics related to our investigation.

\subsection{Regression testing}
Regression testing is a retest activity to ensure that system changes do not affect other parts of the system negatively and that the unchanged parts are still working as they did before a change \cite{minhas2020regression,yoo2012regression}.
It is essential but expensive and challenging testing activity \cite{engstrom2010qualitative}. Various authors have highlighted that testing consumes 50\% of the project cost, and regression testing consumes 80\% of the total testing cost \cite{kazmi2017effective,engstrom2010systematic,engstrom2010qualitative,harrold2008retesting}. Research reports that regression testing may consume more than 33\% of the cumulative software cost \cite{khatibsyarbini2018test}. Regression testing aims to validate that modifications have not affected the previously working code \cite{do2010effects,minhas2020regression}.

Systems and Software Engineering--Vocabulary \cite{ISO/IEC/IEEE8016712}, defines regression testing as:
\begin{displayquote}
\textit{1. ``Selective retesting of a system or component to verify that modifications have not caused unintended effects and that the system or component still complies with its specified requirements.'' \\
2. ``Testing required to determine that	a change to	a system component has not adversely affected functionality, reliability or performance and has not introduced additional defects.''}
\end{displayquote}

For larger systems, it is expensive to execute regression test suites in full \cite{minhas2020regression}. To cope with this, one of the suggested solutions is test case prioritization. It helps to prioritize and run the critical test cases early in the regression testing process. The goal of test case prioritization is to increase the test suite's rate of fault detection \cite{elbaum2002test}. 
 
A reasonable number of systematic literature reviews and mapping studies on various aspects of regression testing provides evidence that regression testing is a well-researched area  \cite{rosero201615, felderer2015systematic, engstrom2010systematic, zarrad2015systematic, kazmi2017effective, catal2012application, yoo2012regression, qiu2014regression, singh2012systematic, catal2013test,bin2019search, khatibsyarbini2018test,bajaj2019systematic,lima2020test,dahiya2018systematic}.
Despite a large number of regression testing techniques proposed in the literature, the adoption of these techniques in the industry is low \cite{rainer2005software, rainer2008follow, engstrom2010qualitative, ekelund2015efficient}. The reasons are that the results of these techniques are not accessible for practitioners due to the discrepancies in terminology between industry and academia \cite{bin2019search, engstrom2010qualitative, minhas2017regression}. There is a lack of mechanisms to guide the practitioners in translating, analyzing, and comparing the regression testing techniques. 
Furthermore, various authors use controlled experiments for their empirical investigations, and in most cases, it is hard to assess that these experiments are repeatable and could fit in an industrial setting \cite{bin2019search}.  Replication of empirical studies could lead us to the desired solution, as it can help to confirm the validity and adaptability of these experiments \cite{shull2008role}.

 \subsection{Replication}

Replication is a means to validate experimental results and examine if the results are reproducible. It can also help to see if the results were produced by chance or the results are the outcome of any feigned act \cite{juristo2012replication}. 
An effectively conducted replication study helps in solidifying and extending knowledge. In principle, replication provides a way forward to create, evolve, break, and replace theoretical paradigms \cite{krein2010case,shull2008role}.
Replication could be of two types 1) internal replication --a replication study carried out by the authors of the original study themselves, 2) external replication --a replication study carried out by researchers other than the authors of the original study \cite{shepperd2018role,krein2010case}.

In software engineering research, the number of internal replications is much higher than external replications \cite{bezerra2015replication,da2014replication}.  Da Silva et al. \cite{da2014replication} reported in their mapping study that out of 133 included replication studies, 55\% of the studies are internal replications, 30\% are external replications, and 15\%  are the mix of internal and external. 
Furthermore, the results of 82\% of the internal replications are confirmatory, and the results of 26\% of external replications conform to the original studies \cite{da2014replication}.
From the empirical software engineering perspective, Shull et al. \cite{shull2008role} classify replications as exact and conceptual replication.  In an exact replication, the replicators closely follow the procedures of the original experiment, whereas, in a conceptual replication, the research questions of the original study are evaluated using a different experimental setup. Concerning exact replication, if the replicators keep the conditions in the replication experiment the same as the actual experiment, it would be categorized as exact dependent replication. If replicators deliberately change the underlying conditions of the original experiment, it would be referred to as exact independent replication. Exact dependent and exact independent replications could respectively be mapped to strict and differentiated replications. A strict replication compels the researchers to replicate a prior study as precisely as possible. In contrast, in a differentiated replication, researchers could intentionally alter the aspects of a previous study to test the limits of the study's conclusions. In most cases, strict replication is used for both internal and external replications \cite{krein2010case}.

\subsection{Information Retrieval}
IR-based techniques are used to retrieve the user's information needs from an unstructured document collection. The information needs are represented as queries \cite{yadla2005tracing,fang2004formal}.
An information retrieval (IR) system is categorized by its retrieval model because its effectiveness and utility are based on the underlying retrieval model \cite{amati2018information}. Therefore, a retrieval model is the core component of any IR system. 

Amati \cite{amati2018information} defines the information retrieval model as: 
\begin{quote}
   \textit{ ``A model of information retrieval (IR) selects and ranks the relevant documents with respect to a user's query. The texts of the documents and the queries are represented in the same way, so that document selection and ranking can be formalized by a matching function that returns a retrieval status value (RSV) for each document in the collection. Most of the IR systems represent document contents by a set of descriptors, called terms, belonging to a vocabulary V.''}
\end{quote}

Some of the retrieval models are the vector space model (VSM), probabilistic relevance framework (PRF), binary independence retrieval (BIR), best match version 25 (BM 25), and language modeling (LM).  
VSM is among the popular models in information retrieval systems. It uses TF-IDF (term frequency and inverse document frequency) as a weighing scheme \cite{roelleke2013information}. 

Since the technique \cite{kwon2014test} we are replicating in this study uses the concepts of TF-IDF weighing scheme, we briefly present TF and IDF. 

Term frequency (TF) and inverse document frequency (IDF) are statistics that indicate the significance of each word in the document or query. TF represents how many times a word appears in the document or query. IDF is an inverse of document frequency (DF). The DF of a word indicates the number of documents in the collection containing the word. Therefore a high IDF score of any word means that the word is relatively unique and it appeared in fewer documents \cite{fang2004formal}.

\section{Related Work}
\label{Sec:RW}

Most of the replication studies on test case prioritization were conducted by the same group of authors, who primarily re-validated/extended the results of their previously conducted experiments (see \cite{do2004empirical,do2006use, do2010effects}).  
Below we discuss studies that are closely related to our topic (i.e., test case prioritization). 

Do et al. \cite{do2004empirical} conducted a replication study to test the effectiveness of the test case prioritization techniques originally proposed for C programs on different Java programs using the JUnit testing framework. The authors' objective was to test whether the techniques proposed for C programs could be generalized to other programming and testing paradigms. The authors who conducted the replication study were part of the original studies, so by definition, it could be referred to as an internal replication. However, concerning the implementation perspective, the replication study would be regarded as differentiated replication.

Do and Rothermel \cite{do2006use} conducted an internal replication study to replicate one of their studies on test case prioritization. The original study used hand-seeded faults. In the replication study, the authors conducted two experiments. In the first experiment, the authors considered mutation faults. The goal was to assess whether prioritization results obtained from hand-seeded faults differ from the results obtained by mutation faults. The authors used the same programs and versions used in the original study. They also replicated the experimental design according to the original study. To further strengthen the findings, later in the second experiment, the authors replicated the first experiment with two additional Java programs with different types of test suites.    

Ouriques et al. \cite{ouriques2018test} conducted an internal replication study of their own experiment concerning the test case prioritization techniques. In the original study, the authors experimented with programs closer to the industrial context. The objective of the replication study was to repeat the conditions evaluated in the original study but with more techniques and industrial systems as objects of study. Although the authors worked with the test case prioritization techniques, they clearly stated that the methods examined in their research use a straightforward operation of adding one test case at a time in the prioritized set. They do not use any data from the test case execution history, and hence, regression test prioritization is not in the scope of their study.

Hasnain et al. \cite{hasnain2019investigating} conducted a replication study to investigate the regression analysis for classification in test case prioritization. The authors' objective to replicate the original study was to confirm whether or not the regression model used in the original study accurately produced the same results as the replicated study. Along with the program and data set used in the original study, the authors also used an additional open-source Java-based program to extend the original study's findings.  It is an external replication study as all authors of the replication study are different from that of the original study. The authors of the replicated study validated the results of the original study on an additional dataset other than the one used in the original study, the replication is not strict.

In the above discussion of related studies, we learned that most replication studies conducted for test case prioritization are primarily internal replications. We could only find a single external replication study \cite{hasnain2019investigating}. The authors of this study conducted the replication of a classification-based test case prioritization using regression analysis. Our study is similar to this study based on the following factors, 1) our study is an external replication, 2) we also use two software artefacts from the original study and two additional artefact. In many ways, our study is unique; for example, 1) we are replicating a technique that focuses on less tested code, whereas Husnain et al. replicated a technique that is based on fault classification and non-faulty modules, 2) we have provided a step by step guide to support future replications, and 3) we provide automated scripts to execute the complete replication study.

\section{Methodology}
\label{Sec:Method}

For reporting the replication steps, we followed the guideline proposal provided by \cite{carver2010towards}. It suggests reporting the following for a replication study:
\begin{enumerate}
    \item Information about the original study (Section \ref{sec:original})
    \item Information about the replication (Section \ref{Sec:Rep})
    \item Comparison of results to the original study (Section \ref{Sec:Res:RQ2})
    \item Drawing conclusions across studies (Section \ref{Sec:con})
\end{enumerate}
\subsection{Research questions}
In the presence of the constraint regarding experimental setup and data, we have to rely on the information presented in the original study (see Section (\ref{sec:original}). We decided not to tweak the original study's approach and followed the steps proposed by the authors and executed the technique on one of the artefacts used by the authors. The differential aspects of the replication experiment are the mutants and the automation of the major steps of the technique. According to the classification provided by Shull et al. \cite{shull2008role}, our work can be classified as \textit{exact independent replication} of the test case prioritization technique presented in \cite{kwon2014test}.

To achieve the objectives of the study we asked the following two research questions:
\begin{displayquote}
\begin{enumerate}
    \item [RQ1.] \textit{To what degree is the study replication feasible given the information provided?} \begin{enumerate}
        \item [RQ1.1] To what degree is the study replicable with the programs used by the original authors?
        \item[RQ1.2] What is the possibility to replicate the study with the new programs?
    \end{enumerate}
    The answer to RQ1 corresponds to Objective 1. While answering RQ1, the motive was to see the possibility to replicate the technique presented in the original study using different programs.

    \item[RQ2.] \textit{Does the replication confirm the findings of the original study?} The answer to RQ2 corresponds to Objective 2. The motive of RQ2 was to see if the replication results conform to the finding of the original study. To ensure that there should be no conscious deviation from the basic technique, we followed the steps and used the tools mentioned in the original study. Finally, we evaluated the replication results using the average percentage of fault detection (APFD) as suggested by the original study's authors. 
\end{enumerate}
\end{displayquote}

\subsection{Information about the original study}
\label{sec:original}

\subsubsection{Selection of target}
Selection of a target study for replication is a difficult process, and often it is prone to biases due to various reasons \cite{pittelkow2021replication}. For example, clinical psychology research reports that authors tend to choose targets that are easy to set up and execute \cite{pittelkow2021replication}. The selection of target must be purpose-based, either by following systematic criteria (see, e.g., \cite{pittelkow2021replication}) or other justifiable reasons. 
In our case, the selection of the target is based on the needs identified from our interaction with industry partners \cite{minhas2017regression,minhas2020regression,bin2019search} and reported facts in the related literature \cite{yoo2012regression, singh2012systematic}.  

For the selection of the target study, our first constraint was test case prioritization, whereas the underlying criteria were to search for a technique that can help control fault slippage and increase the fault detection rate.
During our investigations \cite{minhas2020regression}, we identified that test case prioritization is among the primary challenges for practitioners, and they are interested in finding techniques that can overcome their challenges and help them follow their goals (see also \cite{minhas2017regression}). Increasing the test suite's rate of fault detection is a common goal of regression test prioritization techniques \cite{pan2022test,lima2020test}, whereas controlling fault slip through is among the goals of the practitioners \cite{minhas2017regression,minhas2020regression}. 


Our next constraint was selecting a study where authors used the SIR system to evaluate their technique(s). 
Singh et al. \cite{singh2012systematic} reported that out of 65 papers selected for their systematic reviews on regression test prioritization 50\% are using SIR systems. Yooo et al. \cite{yoo2012regression} also reported that most of the authors evaluate their techniques using SIR artefacts. They highlight that use of SIR systems allows replication studies. 

The final constraint was to select a target study that uses IR methods for the prioritization technique.
Recent studies report that test case prioritization techniques based on IR concepts could perform better than the traditional coverage-based regression test prioritization techniques \cite{peng2020empirically,saha2015information}. 

We searched Google Scholar with the keywords \textit{``regression testing'', ``test case prioritization'', ``information retrieval (IR)'', ``software infrastructure repository (SIR)''}. Our searches returned 340 papers. After scanning the abstracts, we learned that there is not a single technique that explicitly states controlling fault slippage as its goal. However, the technique presented in \cite{kwon2014test} focused on less tested code, and the goal was to increase the fault detection rate of coverage-based techniques using IR methods. Ignored or less tested code could be among the causes of fault slippage. Therefore we considered the technique by Kwon et al. \cite{kwon2014test} for further evaluation. We evaluated this technique using the rigor criteria as suggested by Ivarsson and Gorschek \cite{ivarsson2011method}. The authors suggest evaluating the rigor of empirical studies based on context, design, and validity threats. 

After considering all the factors mentioned above and applying the rigor rubrics, the study presented in \cite{kwon2014test} was used as a target for replication.  

\subsubsection{Describing the original study}

Kwon et al. \cite{kwon2014test} intended to improve the effectiveness of test case prioritization by focusing on infrequently tested code. They argued that test case prioritization techniques based on code coverage might lack fault detection capability. They suggested that using the IR-based technique could help overcome the limitation of coverage-based test case prioritization techniques.  
Considering the frequency at which code elements have been tested,  the technique uses a linear regression model to determine the fault detection capability of the test cases. Similarity score and code coverage information of test cases are used as input for the linear regression model. Kwon et al. \cite{kwon2014test} stated that the technique they proposed is the first of its type that considers less tested code and uses TF-IDF in IR for test case prioritization. The authors claimed that their approach is also first in using linear regression to weigh the significance of each feature regarding fault-finding. They divided the process into three phases, i.e., validation, training, and testing, and suggested using the previous fault detection history or mutation faults as validation and training data.

\vspace{0.2 cm}
 
\begin{center}
        \framebox{

       \parbox[t][5.5cm]{12.0cm}{
       \vspace{0.2 cm}
       \small
     Kwon et al. \cite{kwon2014test} suggested the following steps to implement the proposed technique:
        
    \begin{enumerate}
    \footnotesize{
    \item \textit{Coverage of each test case}
    \item \textit{Set IDF threshold with validation data (previous or mutation faults)}
    \item \textit{Calculate TF/IDF scores of each test case}
    \item \textit{Use coverage and sum of TF/IDF of a test case as predictor values in the training data}
    \item \textit{Use previous (mutation) faults as response values in the training data}
    \item \textit{Estimate the regression coefficients (weight of each feature) with the training data}
    \item \textit{Assign predictor values (coverage and TF/IDF scores) to the model to decide the test schedule}
     \item \textit{Run the scheduled test cases}}
    \end{enumerate}
      
         }
        }
\end{center} 
\vspace{0.2 cm}

To evaluate the proposed test case prioritization technique (IRCOV), \cite{kwon2014test} used four open-source Java programs XML-Security (XSE),  Commons-CLI (CCL), Commons-Collections (CCN), and Joda-Time (JOD). 
\cite{kwon2014test} highlighted that the fault information of the programs was not sufficiently available, and they were unable to evaluate their approach using available information. Therefore, the authors simulated the faults using mutation. To generate the mutants, they used the mutation framework MAJOR \cite{just2011major,just2014major}. To reduce the researcher's bias and achieve reliable results, they applied ten fold validation and divided the mutation faults into ten subsets and assigned each subset to training, validation, and test data.

\subsubsection{Concepts used in the original study}

The original study \cite{kwon2014test} makes use of IR concepts. It views a ``document'' as a test case, ``words'' as elements covered (e.g., branches, lines, and methods), and ``query'' as coverage elements in the updated files. TF and IDF scores of the covered elements determine their significance to a test case. The number of times a test case exercises a code element is counted as a TF value. The document frequency (DF) represents the number of test cases exercising an element. IDF is used to find the unique code elements as it is the inverse of DF. 

Since the focus of the proposed technique was on less-tested code, the IDF score has more significance, and it is required to minimize the impact of TF.  To minimize the impact of TF score on the test case prioritization, they used Boolean values for TF (i.e., $TF = 1$ if a test case covers the code element, $TF = 0$ otherwise). To assign an IDF score to a code element the IDF threshold is used. \cite{kwon2014test} define the IDF threshold as: \\
\begin{displayquote}
\textit{``The maximum number of test cases considered when assigning an IDF score to a code element.''}\\
\end{displayquote}

The IDF threshold is decided by the validation data that consists of faults and related test cases from the previous test execution history or mutation faults.

Finally, the authors used the similarity score between a test case (document) and the changed element (query) to indicate the test cases related to modifications. The similarity score is measured using the sum of TF-IDF scores of common elements in the query.

\subsubsection{Key findings of the original study}

Using four open-source Java programs, the authors compared their technique with random ordering and standard code-coverage-based methods (i.e., line, branch, and method coverage). They measured the effectiveness using Average Parentage of Fault Detection \textit{(APFD)}.

The authors concluded that their technique is more effective as it increased the fault detection rate by 4.7\% compared to random ordering and traditional code coverage-based approaches.

\subsection{Information about the replication}

We first present contextual information, i.e. data availability (Section \ref{SubSec:AC}) and division of the roles during the replication (Section \ref{sec:rolesinvolved}). Thereafter, we describe how the replication steps were implemented (Section \ref{sec:replicationsteps}). 

\subsubsection{Authors' consent and Data availability}
\label{SubSec:AC}
We contacted the original authors to get their consent and ask for any help to replicate their work. We asked them if they can share their experimental package and data with us. We received a reply from one of the corresponding authors and were informed that they do not have any backups related to this study since they conducted this study a few years ago. However, they do not have any objection to the replication of their work.

\subsubsection{Roles involved}
\label{sec:rolesinvolved}

All four authors of this study were given a specified role in the replication. The first and second authors jointly selected the candidate study. The first author conceptualized the whole replication process, including the logical assessment of the study to be replicated. The second author who is an industry practitioner set up the environment according to the requirements of the programs. Both the first and second authors jointly performed the replication steps. The first author provided his input for every step, while the second author carried out the actual implementation. The third and fourth authors reviewed study design and implementation steps.

\subsubsection{Replication steps}
\label{sec:replicationsteps}

\begin{figure}[!htb]
    \centering
    \includegraphics[width=0.5\linewidth]{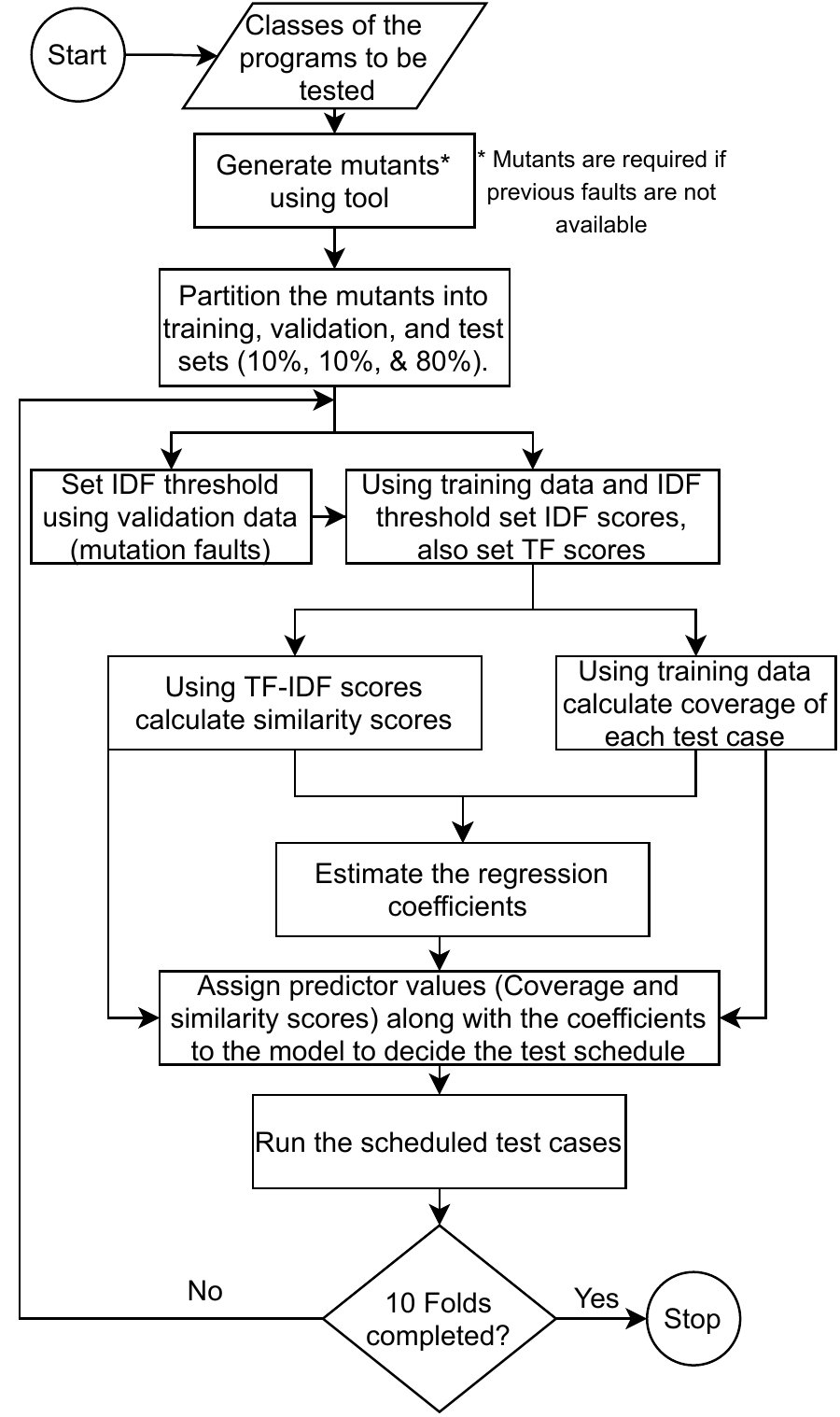}
    \caption{Steps followed to replicate the original study}
    \label{fig:IRCOV}
\end{figure}

We aimed to make an exact replication of the original study, and therefore we followed the procedure strictly as presented in the original study \cite{kwon2014test}. The original study IRCOV was built using total and additional line, branch, and method coverage. However,  \cite{kwon2014test} stated that the results of IRCOV with total and additional coverage were similar. Therefore we only used the total coverage to built the IRCOV models. 
The sequence of events followed in the replication experiment are shown in Figure~\ref{fig:IRCOV}. 

\label{Sec:Rep}

\emph{Replication objects:}
 We built IRCOV models based on three coverage approaches (i.e., Line, Branch, and Method coverage). We aimed to build the IRCOV model using four programs, two from the original study and two new programs. Table \ref{tab:Object} presents the details of the programs used in the replication of IRCOV. The programs are Common CLI, XML security, Commons email, and Log4j. We were able to implement IRCOV with Commons CLI, but due to various constraints discussed in Section \ref{Sec:Res}, we failed to replicate IRCOV with XML security, Commons email, and Log4j. 
\begin{table*}[!htb]
  \caption{Programs used in replication}
     \label{tab:Object}
     \footnotesize
     \centering
     \begin{tabular}{p{2.4cm} c c p{1cm} p{1.5cm} p{2.5cm}} \toprule
       Program & Version & LOC & Test Classes & Used in \cite{kwon2014test}&Repository  \\ \midrule
          Commons CLI & 1.1, 1.2&13210&23& Yes&SIR \& GitHub\\ 
          XML Security&2.2.3&21315&172& Yes&SIR \& GitHub\\ 
          Commons Email& master&83154&20&No& GitHub\\ 
          Log4j & master&169646&63&No&SIR \& GitHub\\ \bottomrule
     \end{tabular}
    
 \end{table*}
 
We selected Commons CLI and XML security as these were used in the original study.  
Commons CLI\footnote{https://commons.apache.org/proper/commons-cli/} is a library providing an API parsing of command-line arguments. 
XML-security\footnote{http://santuario.apache.org/javaindex.html} for Java is a component library implementing XML signature and encryption standards. 
To see if the technique (IRCOV) is replicable with other programs, we selected Commons Email and Log4J. 
Commons Email\footnote{https://commons.apache.org/proper/commons-email/} is built on top of the JavaMail API, and it aims to provide an API for sending email.  

Log4j\footnote{https://logging.apache.org/log4j/2.x/} is a Java based logging utility. Log4j 2 was released in 2014 to overcome the limitations of its predecessor version Log4j 1.  
 We obtained the programs from GitHub and used the test suites provided with the programs.

\emph{Mutant generation:}
The fault information of the programs was not available, and therefore we used mutation faults instead--the authors of the original study used mutation faults. For the mutation, we used the tool (MAJOR) \cite{just2011major,just2014major}. 

\emph{Partitioning mutants into training, validation, and test sets:}
As per the description in the original study, we classified the mutants into training, validation, and test sets (10\%, 10\%, and 80\%, respectively). To classify the data, we used an online random generator\footnote{\label{RGT}https://approsto.com/random-line-picker/}. We applied the ten-fold validation technique to ensure the reliability of the results and avoid any bias. To create ten folds of each data set (i.e., training, validation, and test sets), we wrote automation scripts \cite{replicationStudy}. 

\emph{IDF threshold:}
The purpose of setting up an IDF threshold is to ensure that prioritized test cases should detect faults in less tested code elements. 
The IDF threshold is decided using validation data containing information of faults and of test cases detecting the faults. 
To calculate the IDF threshold the authors of the original study \cite{kwon2014test} suggested using a ratio from 0.1 to 1.0 in Equation \ref{IDFThreshold}.

\begin{equation}
    \label{IDFThreshold}
    IDF\;Threshold = no \;of \;test \;cases \times ratio
\end{equation}
We trained the regression model with each threshold using validation data and selected the ratio that led to the minimum training error for the IDF threshold. Based on the minimum training error, Table \ref{tab:parameters}  presents the chosen values for the IDF threshold of all ten folds of Commons CLI. We assigned IDF values to only those code elements whose DF was not above the IDF threshold.  

\emph{Calculating TF and IDF score:}
As suggested in the original study \cite{kwon2014test}, we use Boolean values for TF (i.e., $TF = 1$ if the test case covers the element, $TF = 0$ otherwise). The purpose to fix the TF values as 0 or 1 was to ensure that only test case would be prioritized that are targeting less tested code. The IDF score is more significant in this regard. As suggested in the original study \cite{kwon2014test}, we used Equation \ref{IDF} to calculate the IDF score. 

\begin{equation}
\label{IDF}
    IDF = 1 + log \left(\frac{\#\;of\;test\;cases}{\#\;of\;test\;cases\;covering\;the\;element}\right)
\end{equation}

\emph{Similarity score:}
The similarity score directly contributes to the IRCOV model. In the regression model (see Equation \ref{IRCOV}), $x\textsubscript{2}$ refers to the similarity score of each test case.
We have calculated the similarity scores using Equation \ref{SScore} as suggested in \cite{kwon2014test}.\\

\begin{equation}
\label{SScore}
    Similarity  \;Score(t,q) = \sum_{e\in t \cap q}tf-idf\textsubscript{e,t}
\end{equation}
Since TF values are 1 or 0 (i.e., if a test case excises a code element, then TF is 1; otherwise, it is 0), practically similarity scores are the sum of IDF scores of the elements covered by a particular test case. 

\emph{Coverage information:}
The coverage measure is aslo used in the regression model. In Equation \ref{IRCOV}, $x\textsubscript{1}$ refers to the coverage size of each test case. To measure code size (line of code) and coverage of each test case, we used JaCoCo\footnote{https://www.eclemma.org/jacoco/}. 

\emph{IRCOV model:}
We used Equation \ref{IRCOV} for the linear regression model as suggested in the original study \cite{kwon2014test}.

\begin{equation}
\label{IRCOV}
    y= \theta\textsubscript{0}+\theta\textsubscript{1}x\textsubscript{1}+\theta\textsubscript{2}x\textsubscript{2}
\end{equation}

In Equation \ref{IRCOV}, $x\textsubscript{1}$ is the size of the coverage data for each test case, and $x\textsubscript{2}$ refers to the similarity score of each test case. The value of y represents each test case's fault detection capability, which is proportional to the number of previous faults detected by the test case. 
In the regression model, three coefficients need to be calculated (i.e., $\theta$\textsubscript{0}, $\theta$\textsubscript{1}, \& $\theta$\textsubscript{2}). Here  $\theta$\textsubscript{0} represents the intersect, whereas, to calculate  $\theta$\textsubscript{1} and $\theta$\textsubscript{2} \cite{kwon2014test} suggested using Equation \ref{theta}, which uses $y$ value and respective values of $x\textsubscript{1}$ and $x\textsubscript{2}$. Here $y$ could be calculated using Equation \ref{Yvalue}, where as $x\textsubscript{1}$ and $x\textsubscript{2}$ respectively represent the size of coverage and similarity scores of each test case.

\begin{equation}
\label{theta}
    theta = (X^T X)^{-1} X^T\vec{y}
\end{equation}

\begin{equation}
\label{Yvalue}
    y= \sum_{n=1}^{n}\frac{f\textsubscript{i}}{log( t\textsubscript{i} ) +1}
\end{equation}

\emph{Prioritization based on fault detection capability:}
After having the values of coefficients and variables of regression model (i.e., $\theta$\textsubscript{0}, $\theta$\textsubscript{1}, $\theta$\textsubscript{2}, $x\textsubscript{1}$, and $x\textsubscript{2}$ ), we determined the fault detection capability of each test case using the IRCOV model (see Equation \ref{IRCOV}). Finally, we arranged the test cases in the descending order of the calculated fault detection capability.  \\

\emph{Evaluating the technique:}
After having a prioritized set of test cases, we ran them on the 50 faulty versions of each fold we created using test data set of mutants. To evaluate the results, we used the average percentage of fault detection (APFD) (see Equation \ref{APFD}). 

\begin{equation}
\label{APFD}
    APFD=1-\frac{TF\textsubscript{1}+TF\textsubscript{2}+....+TF\textsubscript{N}}{nm}+\frac{1}{2n}
\end{equation}

\subsection{Analysis of the replication results}

We implemented IRCOV for line, branch, and method coverage. As described above, for all coverage types, we calculated the APFD values for each fold, we also captured the intermediate results (see Table \ref{tab:parameters}). 

To compare the replication and the original study results, we translated the APFD values for Commons CLI from the original study. Then we plotted the APFD values of the original and replication study in the box plot, a statistical tool to visually summarize and compare the results of two or more groups \cite{williamson1989box, do2010effects}.  Box plot of APFD values enabled us to visually compare the replication and original study results. 

To triangulate our conclusions, we applied hypothesis testing. We used Wilcoxon signed-rank test to compare the results of IRCOV original and IRCOV replication results. Also in the original study \cite{kwon2014test} used Wilcoxon signed-rank test to compare the IRCOV results with the baseline methods. Wilcoxon signed-rank test is suitable for paired samples where data is the outcome of before and after treatment. It measures the difference between the median values of paired samples \cite{gibbons1993location}. In our case, we were interested in measuring the difference between the median APFD values of IRCOV original and IRCOV replication. Therefore, the appropriate choice to test our results was Wilcoxon singed-rank test.

We tested the following hypothesis:
\begin{displayquote}
\begin{enumerate}
    \item[H\textsubscript{0LC}:] There is no significant difference in the median APFD values of original and replication study using line coverage. 
    \item[H\textsubscript{0BC}:] There is no significant difference in the median APFD values of original and replication study using branch coverage.
    \item[H\textsubscript{0MC}:] There is no significant difference in the median APFD values of original and replication study using method coverage.
\end{enumerate}
\end{displayquote}

\subsection{Automation of Replication}
\begin{figure*}
    \centering
    \includegraphics[width=0.6\linewidth]{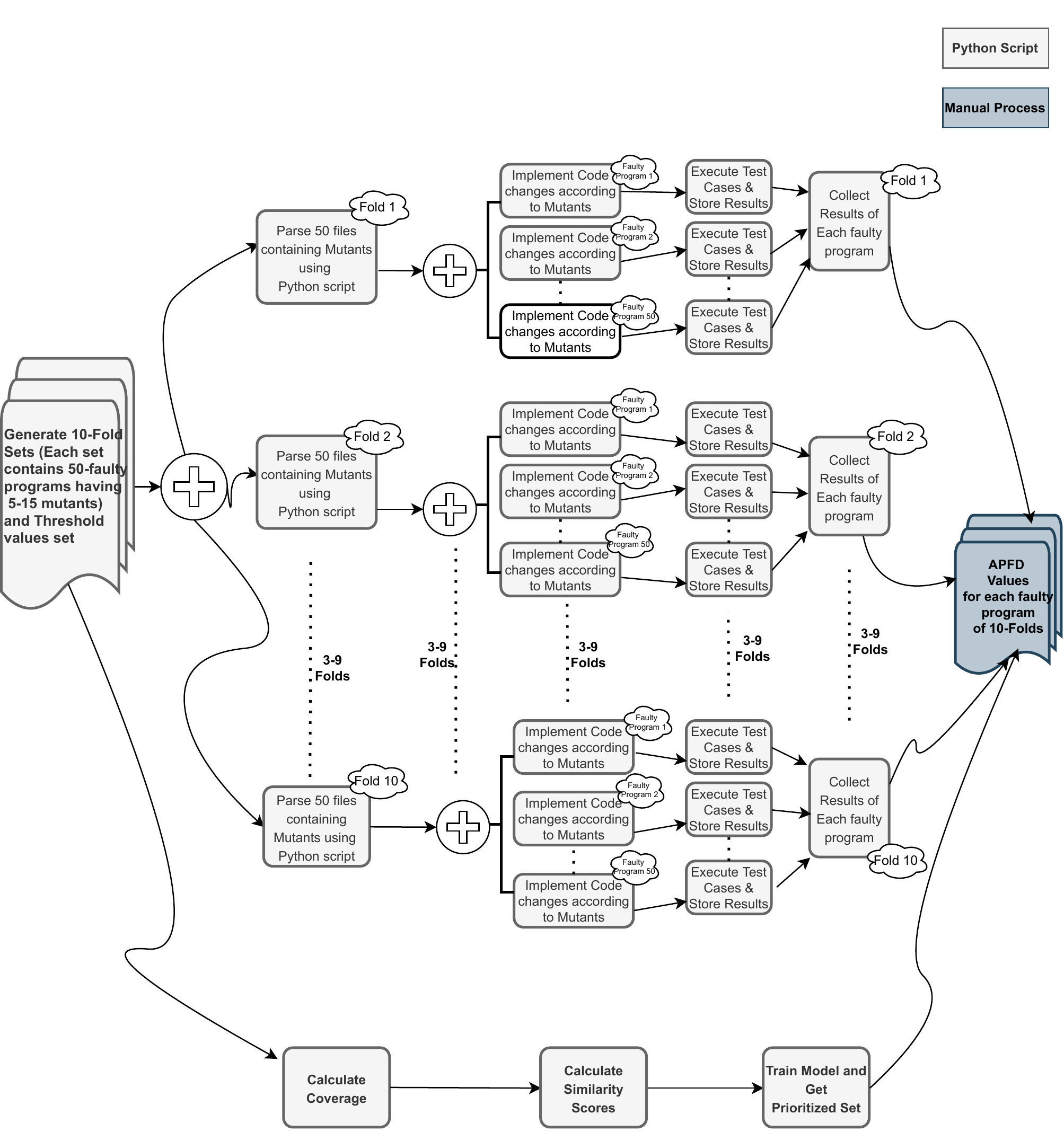}
    \caption{Steps to automate the replication of IRCOV}
    \label{fig:tendFold}
\end{figure*}

The replication was implemented using Python scripts. They are available \cite{replicationStudy}. Figure \ref{fig:tendFold} presents the details of automation steps for the replication of IRCOV.
The original study's authors proposed that ten-fold-based execution is needed (when historical data is not available) to evaluate their original technique. Therefore, our implementation (fold\_generator) \cite{replicationStudy} generates ten folds of the object program at the first stage. Thereafter, it generates fifty faulty versions of each fold, whereas each version contains 5-15 mutants (faults). After generating the faulty versions, the script makes the corresponding changes in the code. Finally, the tests are executed, and their results are extracted. Later, using the test results, we calculate the APFD values of each fold. The calculation of APFD values is the only step not handled in our script. We used excel sheets to calculate APFD values.

\subsection{Threats to validity}
\label{Sec:valid}

\subsubsection{Internal validity}
Internal validity refers to the analysis of causal relations of independent and dependent variables. In our case, we have to see if the different conditions affect the performance of IRCOV. IRCOV depends upon two inputs, coverage of each test case and a similarity score calculated based on TF-IDF. We used the test cases available within the programs. Therefore, we do not have any control over the coverage of these test cases. However, the choices of mutants can impact the similarity score. To avoid any bias, we generated the mutants using a tool and used a random generator to select the mutants for different faulty versions of the programs. Furthermore, we trained IRCOV sufficiently before applying it to test data by following the tenfold validation rule. Since we measured the performance of IRCOV using the APFD measure, the results of the successful case were not significantly different from the original study's results. Therefore we can argue that our treatment did not affect the outcome of IRCOV. Hence minimized the threats to the internal validity. 

\subsubsection{Construct validity}
Construct validity is concerned with the underlying operational measures of the study. In our case, since it is a replication study and we followed the philosophy of exact replication \cite{shull2008role}. Therefore, if the original study suffers of any aspects of construct validity, the replication may do so. For instance, the use of mutation faults could be a potential threat to the construct validity because of the following two reasons
\begin{itemize}
    \item Mutation faults may not be representative of real faults.
    \item Possible researchers' bias concerning the nature of mutation faults. 
\end{itemize}
Concerning the first reason, the use of mutation faults to replace the real faults is an established practice and researchers claim that mutation faults produce reliable results and hence can replace the real faults \cite{do2006use,andrews2005mutation}. 
To avoid any bias, we used an automated mutation tool to generate the mutants. Also to select the mutants for validation, training, and test set we used an automated random selector. Hence no human intervention was made during the whole process.  Furthermore, we discussed the strengths and weaknesses of different tools.

\subsubsection{External validity}
External validity is the ability to ''generalize the results of an experiment to industrial practice'' \cite{wohlin2012experimentation}.  The programs used in the replication study are small and medium-sized Java programs. Therefore, we can not claim the generalizability of results to large-scale industrial projects. The results produced in replication align well with the results of the original study. However, we could not demonstrate the use of the technique on the new programs.

\section{Results}
\label{Sec:Res}
This section presents the findings from the replication. The results are organized according to research questions listed in Section \ref{Sec:Method}.
\subsection{RQ1. Degree to which the replication is feasible to implement.}
\label{Sec:Res:RQ1}
The first goal was to see if it is possible to replicate the IRCOV technique described in the study \cite{kwon2014test}. 

Out of the four replication attempts, we successfully replicated the IRCOV technique with the Commons CLI project. However, with the other three projects (i) XML security, (ii) Commons email, (iii) Log 4j, the replication was either partially successful or unsuccessful due to the reasons elaborated in the following. 

\emph{Successful replication implementation:}
We successfully replicated IRCOV with Commons CLI. After going through the steps presented in Section \ref{Sec:Rep}, for every fold, we were able to calculate the respective coverage information and similarity score of each test case. 
Table \ref{tab:parameters} presents the intermediate results for the replication of IRCOV with Commons CLI. These include, training error, chosen value of IDF threshold, regression coefficient $\theta$\textsubscript{0}, coverage weight $\theta$\textsubscript{1}, and wight for similarity score $\theta$\textsubscript{2}).

   \begin{table*}[!htb]
    \centering
    \caption{Simulation parameters for Commons CLI. (MC = Method coverage, LC = Line coverage, \& BC = Branch coverage)}
    \label{tab:parameters}
    \scriptsize
    \begin{tabular}{p{1cm} p{1.1cm} r c r r r}
    \toprule
 Fold Name &	Coverage Type &	Training Error &	IDF Threshold &	$\theta$\textsubscript{0} &$\theta$\textsubscript{1} &$\theta$\textsubscript{2}\\ \midrule
\multirow{3}{*}{\textbf{Fold1}}
 &	MC&	1.0694	&7&	-0.3478&	0.0187	& 0.1426\\ 

 &	LC	&0.9770 &	2&	0&	0&	0\\ 
  &	BC	&0.8876 &	2&	0 &	0&	0\\ \midrule
 \multirow{3}{*}{\textbf{Fold2}}
 &	MC	&0.3195 &		5 &		-0.7976 &		0.0323 &		-0.1472\\ 
	&	LC&		0.3533 &		6 &		-0.6084&		0.0088&		-0.1343\\ 
	&	BC&		0.3567&		5&		-0.3386&		0.0178&		-0.2095\\ \midrule

\multirow{3}{*}{\textbf{Fold3}}
	&	MC	&	0.6411&		6&		-0.0286&		0.0008 &		0.0796\\ 

	&	LC &		0.6404&		6&		-0.0498&		0.0004 &		0.0736\\ 
	&	BC	&	0.6405 &		6 &		-0.0380 &		0.0008 &		0.0736\\ \midrule
	\multirow{3}{*}{\textbf{Fold4}}
	&	MC &		0.4783 &		6 &		-0.0687 &		0.0097 &		0.1677\\ 
	&	LC	&	 0.4551 &		6 &		-0.1086 &		0.0032 &		0.1365\\ 
	&	BC	&	 0.4947 &		6 &		0.1240 &		0.0045 &		0.1683\\ \midrule
	\multirow{3}{*}{\textbf{Fold5}}
	&	MC	&	0.1838 &		5&		0.0309 &		0.0068&		0.0558\\ 
	&	LC	&	0.1856	&4 &	0.0859 &	0.0018 &		0.0612\\ 
	&	BC	&	0.1876 &		4 &		0.1406 &		0.0038 &		0.0516\\ \midrule
	\multirow{3}{*}{\textbf{Fold6}}
	&	MC	&	0.2247 &		2 &		-0.5284 &		0.0194 &		0.3548\\
	&	LC	&	0.1795 &		2 &		-0.4869 &		0.0052 &		0.3470\\ 
	&	BC	&	0.1549 &		2 &		-0.3978 &		0.0119 &		0.3149\\ \midrule
\multirow{3}{*}{\textbf{Fold7}}
&	MC	&	0.1382 &	10	&	 -0.1479 &		0.0115 &		-0.0141\\ 
	&	LC	&	0.1364	&	 10 &		-0.0833	&	 0.0030 &		-0.0234\\ 
	&	BC	&	0.1390 &		10 &		0.0028 &		0.0065 &		-0.0235\\ \midrule
	\multirow{3}{*}{\textbf{Fold8}}
	&	MC	&	0.2020 &		6 &		0.4389 &		-0.0024 &		0.0839\\ 
	&	LC	&	0.2046 &		6 &		0.3401 &		-0.0001 &		0.0715\\ 
	&	BC	&	0.2046&		6 &		0.3286 &		-0.00001 &		0.0694\\ \midrule
		\multirow{3}{*}{\textbf{Fold9}}
	&	MC	&	0.1490&		6 &		0.1652 &		-0.0032 &		0.1473\\ 
	&	LC	&	0.1532 &		6&		0.0540 &		-0.0002 &		0.1344\\ 
	&	BC	&	0.1517 &		6 &		0.0862 &		-0.0012 &		0.1434\\ \midrule
		\multirow{3}{*}{\textbf{Fold10}}
	&	MC	&	0.0339 &		10 &		-0.1253 &		0.0017 &		0.0267\\ 
&		LC&		0.0339	&	10	&	-0.1127 &	0.0004	&0.0261\\ 
&		BC	&	0.0343 &		10	&	-0.0920	&0.0007&	0.0278\\ \bottomrule

\end{tabular}
\end{table*}

To evaluate the performance of IRCOV, we have calculated APFD values for all ten folds of each coverage type (branch, line, and method) (see Table \ref{tab:apfd}). For branch coverage, the APFD value ranges from 0.547 to 0.873, whereas the average (median) APFD value for branch coverage is 0.747. The APFD values for line coverage range from 0.609 to 0.873, and the average APFD value for line coverage is 0.809. Finally, the APFD value for method coverage ranges from 0.549 to 0.864, and the average APFD for method coverage is 0.772. These results show that the IRCOV model performed best for the line coverage as the mean APFD for line coverage is highest among the coverages.

\emph{Partial or unsuccessful replication:}
Our first unsuccessful replication was concerning XML security. We did not find all the program versions used in the original study (Study \cite{kwon2014test}). Therefore, we decided to use the versions that have slightly similar major/minor release versions. We downloaded available XML security versions 1, 1.5 and 2.2.3. The first two downloaded versions (version 1 and version 1.5) were not compiling due to the unavailability of various dependencies. The logs from the compilation failures are placed in folder ``LogsXmlSecurit'' available at \cite{replicationdata}.

We were able compile the third XML security version 2.2.3, but we could not continue with it, because this version contained several failing test cases (see \cite{replicationdata}). With already failing test cases it was difficult to train the model correctly and get the appropriate list of prioritized test cases. 

The second unsuccessful replication attempt was executed on Commons email. This time the replication was unsuccessful because of faulty mutants generated by the mutant software. For instance, it suggested replacing variable names with 'null' (see Listing 1 \& 2). The actual code was $this.name = null;$ while after mutant injection, the code turned to $this.null = null$.

\begin{lstlisting}[caption=Faulty mutant generated by the tool, basicstyle=\footnotesize]

35:EVR:<IDENTIFIER(java.lang.String)>:<DEFAULT>:org.apache.commons.mail.
ByteArrayDataSource@setName(java.lang.String):214:name |==> null 
\end{lstlisting}

\begin{lstlisting}[caption=Code generated after the insertion of faulty mutant, basicstyle=\footnotesize]

 public void setName(final String name)
    {
                 // this.name = null;  Original code
        this.null =  null;  //Substituted by mutant generator }
\end{lstlisting}

\normalsize
Another type of faulty instances were when MAJOR suggested to modify a line in the code that resulted in Java compilation errors (such as "unreachable statement"). There were several such faulty mutants that made the program fail to compile, and hence no further processing was possible. The detail of all faulty mutants is available in the folder ``CommonsEmail'' at \cite{replicationdata}. 

We also made unsuccessful attempts to change the mutant generator to rectify this problem. However, each mutant generator presented a new set of problems. The lessons learnt from usage of different mutant generators are described in next section.

The third replication attempt was executed on the program Log4j. We followed all the steps (using automatic scripts) proposed by the authors of the original study. We successfully generated the mutants for this program. However, the replication was stopped at the point when the steps to train the model failed. The proposed approach in the original study is based on the coverage information of each code class and test-class. This time the issue was caused by the low coverage of the test cases. During the training of the model, we realized that because of low coverage of the test cases, we were unable to calculate the values of regression coefficients, and as a result, we could not generate the prioritized set of test cases. We developed a Jupyter notebook to describe each steps of this partially successful replication (see \cite{replicationStudy}). Compared to the other programs selected in this study, with 169646 LOC, Log4J is a large program. Thus, a lot of time was needed to train the model for Log4J. For all ten folds, with fifty faulty versions of each fold and with five to fifteen faults in each faulty version, it required approximately 60 hours to train the model.  
  \\

\vspace{0.2 cm}
 
 \begin{center}
        \framebox{ 

       \parbox[t][4.5cm]{12cm}{
       
             \vspace{0.1mm}
     \textbf{Key findings:} Concerning RQ1, the replication was only feasible in one of four cases; the key reasons are listed below.
\begin{enumerate}
\item The inability to use the system under test was caused by compatibility issues (unavailability of system versions and dependencies).
\item Already failing test cases made the replication fail.  
\item Mutant generators created issues in running the replication, workarounds were difficult to implement.
\item Test cases require a certain level of coverage to train the model.
\item More effort is required to train the model for large-sized programs.
\end{enumerate}

         } 
        }
       \end{center} 


\begin{table}[!htb]
    \centering
    \footnotesize
    \caption{APFD values for all ten folds of each coverage type}
    \label{tab:apfd}
    \begin{tabular}{c p{1.2cm} p{1.2cm} p{1.2cm}} \toprule
         Folds&Branch Coverage& Line Coverage& Method Coverage  \\ \midrule
         Fold 1& 0.874&0.874&0.865\\ 
         Fold 2&0.816&0.866&0.790\\ 
         Fold 3&0.646&0.643&0.613\\ 
         Fold 4&0.757&0.816&0.755\\ 
         Fold 5&0.725&0.715&0.721\\ 
         Fold 6&0.796&0.829&0.829\\ 
         Fold 7&0.841&0.839&0.839\\ 
         Fold 8&0.610&0.610&0.585\\ 
         Fold 9&0.548&0.622&0.594\\ 
         Fold 10&0.736&0.803&0.803\\ \bottomrule
    \end{tabular}
    
\end{table}

\subsection{RQ2. Comparison of the results to the original study.}
\label{Sec:Res:RQ2}
\begin{figure}[!htb]
    \centering
    \includegraphics[width=0.6\linewidth]{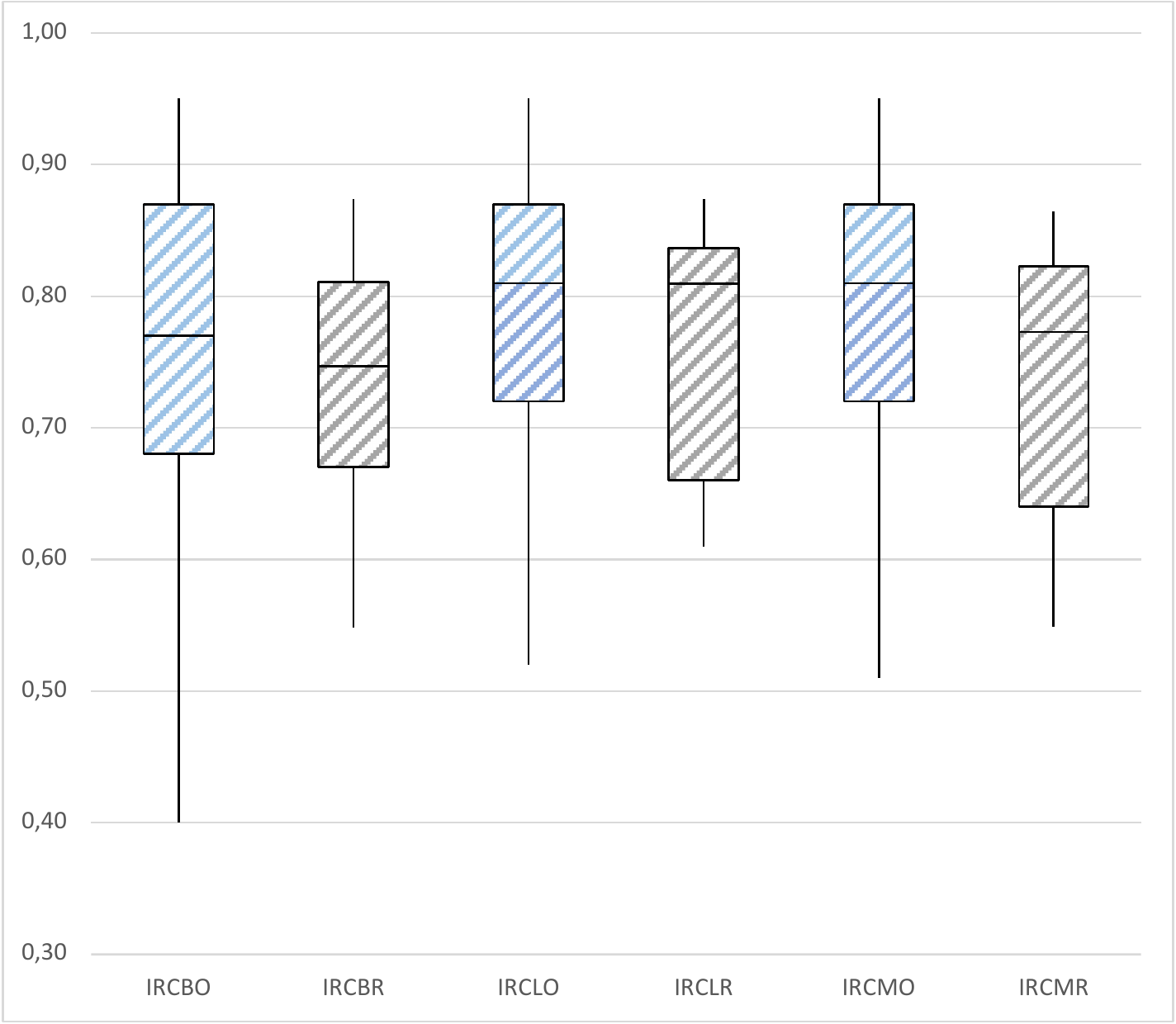}
    \caption{APFD Boxplots for IRCOV Original vs IRCOV Replication\\\\ \small
   IRCBO= IRCOV Branch coverage original, IRCBR= IRCOV Branch Coverage Replication\\
    IRCLO= IRCOV Line coverage original, IRCLR= IRCOV Line coverage replication\\
    IRCMO= IRCOV Method coverage original, IRCMR=IRCOV Method coverage replication}
    \label{fig:APFD_OR}
\end{figure}

Figure \ref{fig:APFD_OR} presents the APFD boxplots of the original and replication study for Commons CLI. Boxplots with blue patterns represent the original study results, and boxplots with gray patterns represent the replication study results. We can see that in all cases, the APFD values of the original study are slightly better compared to the values of the replication. We applied statistical tests to detect whether the results of the replication and the original study differ.

\normalsize
To compare the replication results for branch, line, and method coverage of Commons CLI with the original study's results, we applied Wilcoxon singed-rank test. 
The results are significant if the p-value is less than the level of significance \cite{du2009confidence}.  In our case, the difference between the two implementations would be significant if the p-value is less than 0.05.

Table \ref{tab:stat} presents the results of statistical test. The p-value for branch coverage is 0.475, which is greater than 0.05 (significance level). Therefore, we can not reject the null hypothesis. That means we can not show a significant difference in the APFD values for branch coverage of Commons CLI between the replication and the original study.  

Similarly, the p-value for line coverage is 0.415, greater than the set significance level. Based on the statistical results, we can not reject the null hypothesis. This implies that we can not show a significant difference in the APFD values for the line coverage of Commons CLI between the replication and the original study. 

Finally, the p-value for method coverage is 0.103, based on this result, we can not reject the null hypothesis. Therefore no significant difference in the APFD values for the method coverage of Commons CLI between the replication and the original study.  

\begin{table}[!htb]
    \centering
    \caption{Statistical results of replication compared to the original study for Commons CLI.}
    \begin{tabular}{l c c c} \toprule
        Coverage & $\alpha$ & p-value& 95\% Conf. Int.  \\ \midrule
        Branch &0.05&0.475&  0.646 - 0.816 \\ 
        Line&0.05&0.415&  0.668 - 0.845 \\ 
        Method&0.05&0.103&   0.652 - 0.827\\ \bottomrule
    \end{tabular}
    
    \label{tab:stat}
\end{table}

From the t-test results, we can conclude that for all three coverage types (branch, line, and method), we did not find any significant difference between the replication and the original study. Therefore, we can state that the replication experiment did not deviate from the original result to a degree that would lead to the test detecting a significant difference.

\vspace{0.2 cm}
 
 \begin{center}
        \framebox{

       \parbox[t][4cm]{12cm}{

     \textbf{Key findings:} Concerning RQ2, we compared the replication results of the successful case (i.e., Commons CLI) with the original study's results. Below are the key findings for RQ2.
\begin{enumerate}
    \item The statistical test did not detect a significant difference in the APFD values of the replication and the original study concerning the three coverage measures investigated.
    \item We conclude that the results of the original study are verifiable for Commons CLI. 
\end{enumerate}

         }        }
       \end{center} 


\section{Discussion}
\label{Sec:Disc}

\subsection{Lessons learned of replicating artefact-based studies in software testing}

We replicated the study presented in \cite{kwon2014test} with the intent of promoting artefact-based replication studies in software engineering, validating the correctness of the original study, and exploring the possibilities to adopt regression testing research in the industry.  

Overall, it is essential to capture and document assumptions and constraints concerning the techniques that are replicated, as well as the conditions for being able to run a replication. We highlight several factors of relevance that were observed.

\textit{Conditions concerning System under Test (SuT) complexity:} From the replication results, we learned that besides the various constraints, the technique (IRCOV) presented in \cite{kwon2014test} is replicable for small and medium programs provided the availability of context information. The technique with its current guidelines is difficult to implement with large-size programs because it requires a significant amount of effort to train the model.  For example, the restriction of 10-folds, fifty faulty versions for every fold, and 5 to 15 faults in every faulty version would require a substantial effort (approximately 60 hours) to train the model for large-size programs. This limitation can be managed by reducing the number of faulty versions for each fold, but this may degrade the accuracy and increase the training error.

\textit{Conditions concerning the characteristics of the test suite:} Test cases available with Log4j 2 have low coverage, limiting the chance of correctly training the model and generating a reasonable prioritization order of the test cases. Coverage is one of the primary inputs required for the test case prioritization using the IRCOV model. Another problem we encountered was the presence of already failing test cases in one of the versions of XML security. Test cases are used to calculate the coverage score and similarity scores of the project. If a handful of test cases fail (as in XML security version 2.2.3), wrong coverage information and similarity scores are calculated. This results in the wrong prioritization of test cases as well faulty training of the model (which is used to identify prioritized test cases).
Another drawback with failing test cases concerns the use of mutations. If tests are already failing and when mutants are introduced, then the effectiveness is unreliable as tests are already failing because of other issues. 
Further conditions may be of relevance in studies focusing on different aspects of software testing. Here, we would highlight how important it is to look for these conditions and document them. This is also of relevance for practice, as it demonstrates under which conditions a technique may or may not be successful in practice. 

\textit{Availability of experimental data for artefact-based test replications:} One of the constraints regarding the replicability of the IRCOV technique is the availability of experimental data. For example, the authors of the original study \cite{kwon2014test} stated that they used in-house built tools to conduct the experiment but, they did not provide any source of these tools, also not including details of the automation tools.  Therefore, it took a significant effort to set up the environment to replicate IRCOV with the first program. There are various limitations concerning the data sets and tools required to work with the recommended steps. Regarding data sets, we have recorded the findings in Section \ref{Sec:Res}. These include the compatibility of SIR artefacts. For example, because of various dependencies, we faced difficulties while working with XML security version 1. While working with version 2.2.3 of XML security, we encountered errors in the version. Therefore, we could not collect the coverage information. Ultimately, we were unable to replicate the technique with any of the versions of XML security.

\textit{Reflections on mutant generators:} In the absence of failure data, the authors of the original study suggested using mutation faults, and they used the MAJOR mutation tool to generate the mutants. In one of our cases (Commons Email), the mutation tool (MAJOR) generated inappropriate mutants that led to the build failure. Therefore, no further progress was possible with this case. 

To overcome the difficulty with replication of project 3 (Commons Email), we tried different open-source mutation generators available. Each of these presented various benefits and challenges that are documented in Table  \ref{tab:mutants}. After trying out different mutation tools, we learned that among the available options, MAJOR is an appropriate tool for Java programs, as it generates the mutants dynamically.
\begin{savenotes}

\begin{table*}
    \centering
    \footnotesize
    \caption{Comparison of mutant generators}
    \begin{tabular}{ c p{1.2cm} p{3cm} p{5cm}}
        \toprule
        
No &	Mutation Tool &	Benefits &	Challenges \\   \midrule
1&	Major\footnote{\label{MAJOR}https://mutation-testing.org/}&	(i) Easy to use. (ii) Most commonly used mutant generator. 	&(i) Faulty mutant generated. (ii) Needs upgrade to latest Java versions (iii) Documentation needs improvement.\\  \\ 
2&	$\mu$Java\footnote{https://cs.gmu.edu/~offutt/mujava/}&	(i) IDE plugin available (ii) User decides what types of mutants can be generated.&	(i) Exporting mutants separately is not supported (ii) Does not support latest Java versions (iii) GUI crashes often while generating mutants.\\   \\
3&	Jester\footnote{http://jester.sourceforge.net/}	&Two types of Jester versions, a complete version and a simple version.  &	Latest update is more than 10 years ago. We were unable to generate mutations or start the program despite of following all steps.\\  \\ 
4&	Jumble\footnote{http://jumble.sourceforge.net/}&	(i) Support recent Java versions. (ii) Integration with IDE Supported. 	&Unable to generate mutants despite following examples. Latest update was 6 years ago.\\   \\
5&	PIT\footnote{https://pitest.org/}	&The most recent and complete mutant generator. Mutants are generated and tests are executed. A report is generated for the user. &	(i) Unable to export the mutants. (ii) Lack of diversity in the mutants. (iii) Each execution produced exact same mutants.\\   \bottomrule

    \end{tabular}
    
    \label{tab:mutants}
\end{table*}

\end{savenotes}

\textit{Reflections on the IRCOV technique:} Besides the various limitations highlighted earlier, the IRCOV technique is replicable, and the replication results of the successful case (Commons CLI) show that the original authors' claim regarding the performance of the IRCOV technique was verifiable. The technique presented in the original study can be valuable from the industry perspective because of its focus on prioritizing test cases detecting faults in less tested code while taking coverage of test cases into account during the prioritization process. It can help the practitioners work with one of their goals (i.e., controlled fault slippage). Looking at regression testing in practice, the practitioners recognize and measure the coverage metric \cite{minhas2020regression}. The only information that needs to be maintained in the companies is failure history. In the presence of actual failure data, we do not need to use the mutants to train the IRCOV model extensively, and we can reduce the number of faulty versions for each fold and the number of folds.

Overall, pursuing the first RQ provided us with a deeper insight into the various aspects and challenges related to external replication. The lessons learned in this pursuit are interesting and to provide recommendations in the context of replication studies in software engineering. From the existing literature, it was revealed that the trend of replication studies in software engineering is not encouraging \cite{cruz2019replication, da2014replication}. The studies report that the number of internal replications is much higher than external replications \cite{bezerra2015replication,da2014replication}. While searching the related work, we observed that in the software testing domain, compared to the internal replications, external replications are few in numbers. There could be several reasons for the overall lower number of replication studies in software engineering, but we can reflect on our experiences concerning the external replications as we have undergone an external replication experiment.

One issue we would like to highlight is the substantial effort needed to implement the replication. Replication effort can be substantially reduced with more detailed documentation of the original studies, the availability of appropriate system versions and their dependencies, and the knowledge about prerequisites and assumptions. Better documentation and awareness of conditions may facilitate a higher number of replications in the future.

\subsection{General lessons learned for artefact-based replications}
Table \ref{tab:Recommed} provides an overview of challenges we encountered during the replications. It lists the possible impact of each challenge on the results of replication, and the table also presents a list of recommendations for researchers. The following provides a brief discussion on the lessons learned in this study.  

\begin{table*}[!htb]
    \centering
    \footnotesize
    \caption{Recommendations drawn from the challenges/lessons learned}
    \begin{tabular}{p{3cm} p{3.5cm} p{3.5cm}}
        \toprule
         Challenge& Impact & Recommendation \\   \midrule
         Documentation of original experimental setup& Replicators have to invest additional effort to understand the context of the study.& Original authors need to maintain/publish a comprehensive documentation of experimental setup. \\  \\ 
         Collaboration with the authors of original studies & In the absence of experimental data and support from original authors can make the replication process more complicated. & In the event of request from the replicators the authors of the original study provide assistance in the form of essential information regarding the original experiment. \\  \\ 
         Issues with the open source data sets & Replication experiments may fail due to these issues. & Open source repositories need to maintained and be up to date. \\   \\
        System under Test (SuT) and tools compatibility issues & Any compatibility issue of the tools required to replicate the original experiment can create a bottleneck for the replication. & Such tools (e.g., Mutation tools in our case) need to be maintained to make them compatible with new development frameworks. The same applies to the system under test. \\   \bottomrule
    \end{tabular}
    
    \label{tab:Recommed}
\end{table*}

\emph{Documenting the original experiment:} The authors of the original studies need to maintain and provide comprehensive documentation concerning the actual experiment. The availability of such documents will help the independent replicators understand the original study's context. In the absence of such documentation, the replicators need to invest more effort to understand the original study's context. In this regard, we suggest using open source repositories to store and publish the documentation.  The documentation may contain the detail of the experimental setup, including the tools used to aid the original experiment, automation scripts (if used/developed any), and the internal and final results of the study. Furthermore, the authors can also include detail about any special requirements or considerations that need to be fulfilled for the successful execution of the experiment.

\emph{Collaboration with the original authors:} Because of page limits posed by the journals and conferences, every aspect of the study can not be reported in the research paper. Sometimes, the replicators need assistance from the original authors regarding any missing aspect of the study. Therefore, it is essential that in case of any such query from the replicators, the original study's authors must willingly assist them. Such cooperations can promote replication studies in software engineering. In our opinion, lack of collaboration is one reason for fewer replication studies in software engineering. However, it is important to still conduct the replications as independently as possible due to possible biases (i.e., avoiding to turn an external replication into an internal one).

\emph{Maintaining open source repositories:} Open-source repositories (one example being SIR) provide an excellent opportunity for researchers to use the data sets while conducting software engineering experiments. A large number of researchers have benefited from these data sets. We learned that some of the data sets available in repositories are outdated and need to be maintained. Such data sets are not helpful, and studies conducted using these data sets would be complex to adopt/replicate. It is therefore essential that authors explicitly state the the versions they used in their own studies. In addition, we recommend that authors of original studies as well as replications ensure that the dependencies or external libraries are stored to avoid that the system under test can not be used in replications.

\emph{Tools compatibility:} In many cases, the authors need to use open source tools to assist the execution of their experiment. Such tools need to be well maintained and updated. In case of compatibility issues, these tools can hinder the replication process. For example, the study we replicated uses a mutation tool (MAJOR). Although it is one of the best choices among the available options, the tool generated inappropriate mutants for one of our cases due to some compatibility issues. Ultimately, after a significant effort, we had to abandon the replication process for that case. Here, we also would like to highlight that one should document the versions of the tools and libraries used (also including scripts written by the researchers - e.g., in Phython). 

\emph{Documenting successes and failures in replications:} Besides the significance of documenting every aspect of the original experiment, recording every single event of replication (success \& failure) is critical for promoting future replications and industry adoptions of research. We recommend storing the replication setups and data in open source repositories and providing the relevant links in the published versions of the articles. 

\emph{Automation of replication:}
A key lesson learned during the replication of the original study is that the documentation of the setup and execution of replication could be automated with the help of modern tools and programming languages.  This automation will help in reproducing the original results and analysis for researchers reviewing or producing the results from the studies. We have provided programming scripts that describe and documented all the steps (and the consequences of these steps). 

\section{Conclusions}
\label{Sec:con}
This article reports the results of a replication of the test case prioritization technique using information retrieval (IR) concepts proposed initially by \cite{kwon2014test}. We replicated the original study using four Java programs: Commons CLI, XML security, Commons email, and Log4j. We selected two programs from the original study, and the other two were new.   
We aimed to answer the two research questions. In RQ1, the aim was to see if the technique is replicable, and in RQ2, we aimed to see if the replication results conform to the ones presented in the original study. 

We have faced various challenges while pursing RQ1, these challenges are related to the availability of original experimental setup, collaboration with the original authors, system under test, test suites, and compatibility of support tools. We learned that the technique is replicable for small programs subject to the availability of context information. However, it is hard to implement the technique with the larger programs because it requires a substantial effort to train it for a larger program.

To verify the original study's results (RQ2), we compared the replication results for Commons CLI with the ones presented in the original study. These results validated the original study's findings as the statistical test confirms no significant difference between the APFD values of the replication and the actual experiment. However, we must say that our results partially conformed with the original study because we could not replicate the technique with all selected artefacts due to missing dependencies, broken test suites, and other reasons highlighted earlier.

The technique can be helpful in the industry context as it prioritizes the test cases that target the less tested code. It can help the practitioners to control fault slippage. However, it needs some improvements in training and validation aspects to scale the technique to the industry context. To support the future replications/adoption of IRCOV, we have automated the IRCOV steps using Python (Jupyter notebook). 

We plan to work with more artefacts with actual faults to test the technique's (IRCOV) effectiveness in the future, and we plan to see the possibilities of scaling it up for larger projects.
In addition to that, we want to evaluate our proposed guidelines (under lessons learned) using different studies from industrial contexts.

\small
\vspace{\baselineskip}
\noindent
\textbf{Author contributions}
All authors have contributed to every phase of this study, i.e., the conception of the idea, implementation, and manuscript writing.
All authors read and approved the final manuscript, and they stand accountable for all aspects of this work's originality and integrity.

\section*{Acknowledgement}
This work has in parts been supported by ELLIIT; the Swedish Strategic
Research Area in IT and Mobile Communications.

\normalsize
\bibliographystyle{abbrv}      

\bibliography{IRCOV}   

\end{document}